\documentstyle[prl,aps,epsfig]{revtex}

\widetext

\draft

\begin{document}

\title{\bf  Non-dissipative decoherence bounds on quantum computation} 

\author{Stefano Mancini and Rodolfo Bonifacio}

\address{INFM, Dipartimento di Fisica,
Universit\`a di Milano, Via Celoria 16, I-20133 Milano, Italy}

\date{\today}

\maketitle

\begin{abstract}
We investigate the capabilities of a quantum computer based on cold 
trapped ions in presence of
non-dissipative decoherence. The latter is accounted by using the 
evolution time as a random
variable and then averaging on a properly defined probability 
distribution. 
Severe bounds on computational performances are found.
\end{abstract}


\section{Introduction} 

A quantum computer processes quantum information that is stored 
in quantum bits (qubits)
\cite{STEANE}. If a small set of fundamental operations, 
or ``universal quantum logic gates" can be
performed on the qubits, then a quantum computer can be 
programmed to solve an arbitrary problem
\cite{DEU}. Essentially, the quantum computation can be 
viewed as a coherent superposition of
digital computation proceeding in parallel. The explosion 
of interest in quantum computation can be
traced to Shor's demonstration that a quantum computer could 
efficiently factorize large integers
\cite{SHOR}. 

Cirac and Zoller \cite{CIZO} have made one of the most promising 
proposal for the
implementation of a quantum computer. A number of identical 
atoms are stored and laser cooled in a
linear radio-frequency quadrupole trap to form a quantum register. 
The radio-frequency trap
potential gives strong confinement of the ions in the directions 
transverse to the trap axis, while
an electrostatic potential forces the ions to oscillate in an 
effective harmonic potential in the
axial direction. After laser cooling the ions become localized along 
the trap axis with a spacing
determined by their Coulomb repulsion and the confining 
axial potential.The normal mode of the ions'
collective oscillations which has the lowest frequency, is the 
axial center of mass (CM) mode, in
which all the trapped ions oscillate together. A qubit is the 
electronic ground state $|g\rangle$
and a long-lived excited state $|e\rangle$ of the trapped ions. 
The electronic configuration of
individual ions, and the quantum state of their collective CM 
vibrations can be manipulated by
coherent interactions of the ion with a laser beam, in a standing 
wave configuration, which can be
pointed at any of the ions. The CM mode of the axial vibrations may 
then be used as a ``bus" to
implement the quantum logical gates. Once the quantum computation 
has been completed, the readout is
performed through the mechanism of quantum jumps. Several features 
of this scheme have been
demonstrated experimentally, mostly  using a single 
trapped ion \cite{EXP}. 

The implementation
of a large-scale quantum computer is recognized to be a 
technological challenge of unprecedent
proportions. In fact the qubits must be easily manipulated, 
but must be also well isolated from
decohering influence which places considerable limitations 
on the capabilities of such devices
\cite{LAND}. Practically, there are two fundamentally 
different types of decoherence during a
computation: the dissipative one, due to the spontaneous 
decay of the metastable states $|e\rangle$,
and the non-dissipative one due to random phase fluctuations 
of various nature. While the first has
been considered carefully \cite{PLENIO}, by also developing 
strategies to reduce spontaneous decay,
like watchdog stabilization \cite{ZUR84}, the second does not 
received much attention Ref.
\cite{ZUR}.  

On the other hand, non-dissipative decoherence seems actually 
dominant in trapped
ion based experiments \cite{WIN}, whose results have been 
recently well explained \cite{BOTO}, both
qualitatively and quantitatively, by simply considering the 
time as a statistical variable and then
averaging over a properly defined probability distribution 
\cite{RB}. The developed approach results
model-independent, hence quite ductile. Then, our aim here, 
is to study the performances of an
ion-trap quantum information processor by using this approach. 
Practically, the latter allows us to account for the 
fluctuations of the Raman laser intensity
\cite{NIST}. Instead, in \cite{ZUR} a specific model 
is used, i.e. a random phase drift is added by
hand at the end of each qubit rotation while in our 
case the stochasticity is already included in
the time evolution. Moreover, our analysis is
based on  exact parameters values fitting experimental 
data \cite{BOTO}.

The paper is organized as follows.
In Section II we shall review a general model for 
non-dissipative decoherence. In Section III we
shall apply this theory to fundamental logic operations. 
In Section IV we evaluate the limits of a
quantum computer due to this type of decoherence. 
Finally, Section V is for concluding remarks.

\section{A general formalism for non-dissipative decoherence}

In this section we
review the formalism describing non-dissipative decoherence derived 
in Refs. \cite{RB}. It is based
on the idea that time is a random variable or, alternatively, 
that the system Hamiltonian (therefore
its eigenvalues) fluctuates. 
This leads to random phases in the energy eigenstates representation.
Then, the resulting evolution of the system must be averaged on a 
suitable probability 
distribution, and this leads to the decay of the off-diagonal 
elements of the density operator. 

Let us consider an initial state $\rho(0)$ and consider the case 
of a random evolution time. Then,
the evolved state will be averaged over a probability distribution 
$P(t,t')$, i.e.
\begin{equation}\label{rhobardef}
{\overline\rho}(t)=\int^{\infty}_0\, dt' \, P(t,t')\, \rho(t')\,,
\end{equation}
where $\rho(t')=\exp\{-iLt'\}\rho(0)$ is the usual solution of 
the Liouville-Von Neumann equation
with $L\ldots=[H,\ldots]/\hbar$. One can write as well 
\begin{equation}\label{rhobarV}
{\overline\rho}(t)=V_L(t)\rho(0)\,, 
\end{equation} where  
\begin{equation}\label{VP} 
V_L(t)=\int^{\infty}_0\, dt' \, P(t,t')\,e^{-iLt'}\,.  
\end{equation} 
In Ref. \cite{RB}, the
function $P(t,t')$ has been determined 
to satisfy the following conditions:
i) ${\overline\rho}(t)$
must be a density operator, i.e. it must be self-adjoint, 
positive-definite, and with unit-trace.
This leads to the condition that $P(t,t')$ must be 
non-negative and normalized, i.e. a probability
density in $t'$, so that Eq.(\ref{rhobardef}) is a 
completely positive map; ii) $V_L(t)$ satisfies
the semigroup property $V_L(t_1+t_2)=V_L(t_1)V_L(t_2)$, 
with $t_1, t_2 \ge 0$. These requirements
are satisfied by 
\begin{equation}\label{V} V_L(t)=\frac{1}{(1+iL\tau)^{t/\tau}}\,,  
\end{equation} 
and  
\begin{equation}\label{P} 
P(t,t')=\frac{1}{\tau}\frac{e^{-t'/\tau}}{\Gamma(t/\tau)}
\left(\frac{t'}{\tau}\right)^{(t/\tau)-1}\,, 
\end{equation} 
where the parameter $\tau$ naturally appears as a scaling time. 
The expression
(\ref{P}) is the so-called $\Gamma$-distribution function, well 
known in line theory. Its
interpretation is  particularly simple when $t/\tau=k$ with $k$ 
integer; in that case $P(k,t')$
gives the probability density that the waiting time for $k$ 
independent events is $t'$ and $\tau$ is
the mean time interval between two events. Generally, the 
meaning of the parameter $\tau$ can be
understood by considering the mean of the evolution 
time $\langle t'\rangle=t$, and its variance
$\langle t'^2\rangle-\langle t'\rangle^2=\tau t$.  

When $\tau\to 0$, $P(t,t')\to \delta(t-t')$ so
that ${\overline\rho}(t)\equiv\rho(t)$ and $V_L(t)=\exp\{-iLt\}$ 
is the usual unitary evolution.
However, for finite $\tau$, the evolution operator $V_L(t)$ 
describes a decay of the off diagonal
matrix elements in the energy representation, whereas the 
diagonal matrix elements remain constants,
i.e. the energy is still a constant of motion. In fact, 
in the energy eigenbasis,
Eqs.(\ref{rhobarV}) and (\ref{V}) yield  
\begin{equation}\label{rhobarnm}
{\overline\rho}_{n,m}(t)=e^{-\gamma_{n,m}t}\,e^{-i\nu_{n,m}t}
\,\rho_{n,m}(0)\,, 
\end{equation}
where 
\begin{eqnarray}
\gamma_{n,m}&=&\frac{1}{2\tau}
\log\left(1+\omega_{n,m}^2\tau\right)\,,\label{ga}
\\
\nu_{n,m}&=&\frac{1}{\tau}\arctan\left(\omega_{n,m}\tau\right)
\,,\label{nu}
\end{eqnarray} 
with
$\hbar\omega_{n,m}=(E_n-E_m)$ the energy difference. 
One can recognize in Eq.(\ref{rhobarnm}),
beside the exponential decay, a frequency shift of every 
oscillating term. 

The phase diffusion
aspect of the present approach can also be seen in 
the evolution equation for the averaged density
matrix ${\overline\rho}(t)$. Indeed, by differentiating 
with respect to time Eq.(\ref{rhobarV}) and
using (\ref{V}) one gets the following master equation 
for ${\overline\rho}(t)$
\begin{equation}\label{meq} 
\frac{d}{dt}{\overline\rho}(t)= -\frac{1}{\tau}\log
\left(1+iL\tau\right) {\overline\rho}(t)\,. 
\end{equation} 
It is worth noting that by expanding
the logarithm up to second order in $\tau$, one obtains 
\begin{equation}\label{meqapp}
\frac{d}{dt}{\overline\rho}(t)=
-\frac{i}{\hbar}\left[H,{\overline\rho}(t)\right]
-\frac{\tau}{2\hbar^2}\left[H,\left[H,{\overline\rho}(t)
\right]\right]\,,
\end{equation}
which is the well known phase-destroying master equation \cite{GAR}.
Hence, Eq.(\ref{meq}) appears as a generalized phase destroying 
master equation taking into account
higher order terms in $\tau$. Nonetheless, the present approach 
is different from the
usual master equation approach, in the sense that it is model 
independent, non perturbative 
and without specific statistical assumptions.

The theory well describes also Hamiltonian fluctuations. 
Let us consider, for instance, a Hamiltonian 
$H(t)=\Omega(t)\tilde{H}$,
where $\Omega(t)$ is a fluctuating parameter
with mean value $\Omega$.
Whenever $\tilde{H}=H/\Omega$ induces oscillations,
we can accounts for the fluctuations of the frequency $\Omega(t)$
by writing \cite{BOTO} 
\begin{equation}\label{rhoA}
\rho(t)
=\exp\left\{-i\tilde{L}A(t)\right\}\rho(0)\,,
\quad
\Longrightarrow
\quad
\overline{\rho}(t)=\int_0^{\infty}\,
dA\,P(t,A)e^{-i\tilde{L}A}\rho(0)\,,
\end{equation}
where $\tilde{L}=[\tilde{H},\ldots]/\hbar$,
and $A(t)=\int_0^t \, d\xi \, \Omega(\xi)$
is a positive dimensionless random variable.
Furthermore,
\begin{equation}\label{PA}
P(t,A)=\frac{1}{\Omega\tau}\frac{e^{-A/\Omega\tau}}{\Gamma(t/\tau)}
\left(\frac{A}{\Omega\tau}\right)^{(t/\tau)-1}\,.
\end{equation} 
The first two moments of $P(t,A)$ determine the properties of the 
fluctuating frequency $\Omega(t)$,
\begin{equation}\label{statA}
\langle A \rangle =\Omega t\,,
\quad
\sigma^2(A)=\langle A^2 \rangle-\langle A \rangle^2= 
\Omega^2 t \tau\,,
\end{equation}
that is, the frequency $\Omega(t)$ is a white, non-gaussian 
(due to the non-gaussian form of $P(t,A)$) stochastic process.
In fact, the semigroup assumption made, implies a Markovian 
treatment in which the spectrum of the fluctuations 
is flat in the 
relevant frequency range. This in particular implies 
that we are neglecting 
the dynamics at small times, of the order of the 
correlation time of the fluctuations.
The non-gaussian character of $P(t,A)$ can be 
traced back to the fact 
that $P(t,A)$ must be definite and normalized in the interval 
$0<A<\infty$  and not in $-\infty <A< \infty$.
This is another fundamental difference 
with respect to Ref. \cite{ZUR}, 
where a gaussian
character for the random phase drift errors
is assumed, while we make
no a priori statistical assumptions.
Indeed, the properties of the probability distribution $P$ 
are derived only from the semigroup
condition, and it is interesting to note that this condition 
yields a gaussian probability
distribution  as a limiting case. In fact, from Eq.(\ref{P}) 
one can see that $P(t,t')$ tends to
become a gaussian  in the large time limit $t\gg\tau$.

In the case of Rabi oscillations, $A$ becomes 
proportional to the pulse area, 
while $\tau$ gives an estimate of the pulse 
area fluctuations, since it corresponds to a 
fractional error of the pulse 
area $\sqrt{\sigma^2(A)}/\langle A\rangle=\sqrt{\tau/t}$.

\section{Characterization of a quantum process}

Quantum computation ideally corresponds to a physical process 
$|\Psi_{\rm out}\rangle= U |\Psi_{\rm in}\rangle$ where a given input 
state is mapped
to an output state by a unitary transformation $U$.
It has been shown that any quantum computation can be decomposed into 
one-bit gates and 
a universal two-bit gate which involves entanglement operation on two 
qubits \cite{EKJO}.

In order to characterize a physical process in a quantum system, 
we follow
Ref. \cite{POYATOS}. 
The aim is to characterize the process given as a ``black box", by 
a sequence of measurements 
in such a way that it is possible to predict what the output state 
will be for any input state.

We assume that the system is initially prepared in a pure state
\begin{equation}\label{rhoin}
|\Psi_{\rm in}\rangle=\sum_{i=0}^N c_i \, |i\rangle\,,
\quad\Rightarrow\quad
\rho_{\rm in}=|\Psi_{\rm in}\rangle\langle\Psi_{\rm in}|\,,
\end{equation}
where $|0\rangle,\,|1\rangle\ldots,|N\rangle$ are orthogonal 
states spanning the Hilbert space 
of allowed input states with dimension $N+1$. 
Then, in the ideal case, the output state can be written as
\begin{equation}\label{rhoout}
|\Psi_{\rm out}\rangle\langle\Psi_{\rm out}|\equiv\rho_{\rm out}
=\sum_{i,i'=0}^N c_i c_{i'}^* \,
R_{i',i}\,, 
\end{equation}
where 
\begin{equation}\label{R}
R_{i',i}=U |i\rangle\langle i'| U^{\dag}
\,,
\end{equation}
are system operators not depending on the input state. 

Whenever the unitary transformation is given by the time 
evolution operator $U(t)$, the initial state $\rho(0)$ 
represents the 
input state while the evolved state $\rho(t)$ represents the
output one.
Since in Section II  $\rho(t)$ is replaced by ${\overline\rho}(t)$,
here we have to replace $R_{i',i}(t)$ by
\begin{equation}\label{Rbar}
{\overline R}_{i',i}(t)=\int_0^{\infty}\, dt'\, P(t,t') \,
U(t') |i\rangle\langle i'| U^{\dag}(t')
\,.
\end{equation}

Now, in order to see to what extent the real physical process 
approaches the ideal one,
we use as parameter the fidelity
\begin{equation}\label{Fdef}
{\cal F}={\rm Tr}\left\{\rho(t){\overline\rho}(t)
\right\}_{\rm ave}\,,
\end{equation}
where the subscript ''${\rm ave}$" indicates the average overall 
possible input states.
Obviously a gate fidelity close to one indicates that the gate 
was carried out almost ideally. 

In what follows we will see that the fidelity
${\cal F}$ (\ref{Fdef}) can be expressed in terms of the tensor
\begin{equation}\label{Ftens}
F^{i'\,i}_{j'\,j}\equiv\langle j'|U^{\dag} 
{\overline R}_{i'\,i} U|j\rangle\,.
\end{equation}

\subsection{One-bit gate}

In the case of one bit $N=1$, and the vector basis is given by 
$\{ |i=0\rangle\equiv|g\rangle,\, 
|i=1\rangle\equiv|e\rangle \}$.
The individual rotation acting on a single ion can be performed 
using a laser frequency on resonance
with the internal transition and with  the equilibrium position 
of the ion coinciding with the
antinode of the laser standing wave. The Hamiltonian describing 
such a interaction is \cite{CIZO}
\begin{equation}\label{Hone}
H=\frac{\Omega}{2}\left[ |e\rangle\langle g| e^{-i\phi}
+ |g\rangle\langle e| e^{i\phi}\right]\,,
\end{equation}
where $\Omega$ is the Rabi frequency and $\phi$ the laser phase.
Eq.(\ref{Hone}) leads to the following evolution 
\begin{eqnarray}
U(t)|g\rangle&=&
\cos\left[\Omega t/2\right]|g\rangle
-ie^{i\phi}\sin\left[\Omega t/2\right]
|e\rangle\,,\label{Ug}
\\
U(t)|e\rangle&=&\cos\left[\Omega t/2
\right]|e\rangle
-ie^{-i\phi}\sin\left[\Omega t/2\right]
|g\rangle\,,\label{Ue}
\end{eqnarray}
which effectively corresponds to a single qubit rotation.
It is clear from the above evolution equations that 
fluctuations in the Rabi frequency can be
accounted by simply replacing $\Omega t$ with
$A(t)=\int_{0}^t \, d\xi \, \Omega(\xi)$ (where now $\Omega$ 
indicates the mean Rabi frequency) accordingly to the arguments of
Section II.
Then, similarly to Eq.(\ref{Rbar}), it is possible to
evaluate
\begin{equation}
{\overline R}_{i',i}(t)=\int_0^{\infty}\, dA\, P(t,A) \,
U(A) |i\rangle\langle i'| U^{\dag}(A)
\,,
\end{equation}
and the tensor elements (\ref{Ftens}) (see Appendix I) .
In terms of these elements the fidelity results
\begin{equation}\label{Fone}
{\cal F}=\frac{3}{8}\sum_{i=0}^1 F^{i\,i}_{i\,i}+
\frac{1}{8}\sum_{i \ne j=0}^1 \left(F^{i\,i}_{j\,j}
+F^{j\,i}_{i\,j}\right)
\,.
\end{equation}
In this case the fidelity depends on time which 
practically determines the amount 
of rotation, however, in the next Section we shall 
consider a $\pi$ rotation, i.e. $t=\pi/\Omega$.

\subsection{Two-bit gate}

In the case of two-bit $N=3$,  and the vector basis is given by
$\{|i=0\rangle\equiv|g\rangle_1|g\rangle_2\,, 
|i=1\rangle\equiv|g\rangle_1|e\rangle_2,\,
|i=2\rangle\equiv|e\rangle_1|g\rangle_2,\,
|i=3\rangle\equiv|e\rangle_1|e\rangle_2\}$,
where labels 1, 2 indicate the two bits. 
Beside that also the vibrational ground state $|0\rangle$ is 
employed.

In this case the laser frequency is chosen tuned to the first
motional sideband and the equilibrium position of the ion 
coincides with the node of the laser
standing wave.  
The Hamiltonian describing such interaction is \cite{CIZO}
\begin{eqnarray}\label{Htwo}
H_{n,q}=
\left\{
\begin{array}{l}
\frac{\Omega'}{2}
\left[ |e\rangle_n\langle g| a e^{-i\phi}
+ |g \rangle_n\langle e| a^{\dag} e^{i\phi}\right]\,,
\quad q=0\,,
\\
\\
\frac{\Omega'}{2}
\left[ |e' \rangle_n\langle g| a e^{-i\phi}
+ |g \rangle_n\langle e'| a^{\dag} e^{i\phi}\right]\,,
\quad q=1\,,
\end{array}
\right.
\quad
\Omega'=\frac{\eta\Omega}{\sqrt{N_a}}\,.
\end{eqnarray}
Here, $a^{\dag}$ and $a$ are the creation and annihilation 
operators of the CM phonons,
$\Omega$ is the Rabi frequency, $\phi$ is the laser phase, 
and $\eta$ is the Lamb-Dicke 
parameter. The index $q=0,1$ refers to the transitions 
excited by the laser,
$|g\rangle\leftrightarrow |e\rangle$ or 
$|g\rangle\leftrightarrow |e'\rangle$,
which depend on the laser polarization.
Instead the index $n$ refers to the $n$-th ion on the 
trap (1, 2 in our case).
Moreover, the factor $\sqrt{N_a}$,
where $N_a$ indicates the number of trapped ions, appears 
as a consequence of the 
M\"ossbauer effect \cite{CIZO}. 

The universal two-bit gate, defined by \cite{EKJO}
\begin{equation}\label{univ}
|\epsilon_1\rangle_1\,|\epsilon_2\rangle_2
\to (-1)^{\epsilon_1 \epsilon_2}
|\epsilon_1\rangle_1\,|\epsilon_2\rangle_2\,,
\quad 
(\epsilon_{1,2}=0,1)
\end{equation}
can be realized in three steps by means of Eq.(\ref{Htwo}):
\begin{itemize}
\item{
A $\pi$ laser pulse with polarization $q=0$ and $\phi=0$ 
excites e.g. the first ion.
The evolution will be
\begin{equation}\label{UI}
U_I(t_1)=\exp\left[-i\frac{\Omega'}{2}
t_1\left(|e\rangle_1\langle g| a+|g\rangle_1\langle
e| a^{\dag}\right)
\right]\,, 
\end{equation}
with $t_1=\pi/\Omega'$.
}
\item{
The laser directed on the second ion is then turned 
on for a time of a $2\pi$
pulse with polarization $q=1$ and $\phi=0$.
The evolution will be
\begin{equation}\label{UII}
U_{II}(t_2)=\exp\left[-i\frac{\Omega'}{2}
t_2\left(|e'\rangle_2\langle g| a+|g\rangle_2\langle
e'| a^{\dag}\right) \right]\,, 
\end{equation}
with $t_2=2\pi/\Omega'$.
}
\item{
A $\pi$ laser pulse with polarization $q=0$ and $\phi=0$ 
excites again the first ion.
The evolution will be
\begin{equation}\label{UIII}
U_{III}(t_3)=\exp\left[-i\frac{\Omega'}{2}
t_3\left(|e\rangle_1\langle g| a+|g\rangle_1\langle
e| a^{\dag}\right) \right]\,, 
\end{equation}
with $t_3=\pi/\Omega'$.
}
\end{itemize}
Also in this case to account for
the fluctuations in the Rabi frequency,
we can introduce a stochastic variable 
$A'(t)=\int_{0}^t \, d\xi \, \Omega'(\xi)
=\int_{0}^t \, d\xi \, \Omega(\xi)\eta/\sqrt{N_a}$
(where now $\Omega'$ is related to the mean Rabi frequency $\Omega$ 
as in (\ref{Htwo})).
Then, we have
\begin{eqnarray}\label{Rbartwo}
{\overline R}_{i'\,i}&=&
\int_0^{\infty}\, dA_3'\,
\int_0^{\infty}\, dA_2'\,
\int_0^{\infty}\, dA_1'\, P(t_3,A_3') P(t_2,A_2') P(t_1,A_1') 
\nonumber\\
&&\times
U_{III}(A_3') U_{II}(A_2') U_{I}(A_1') |i\rangle
\langle i'| U_{I}^{\dag}(A_1') U_{II}^{\dag}(A_2') 
U_{III}^{\dag}(A_3')
\,,
\end{eqnarray}
which are used to calculate the tensor elements (\ref{Ftens})
(see Appendix II).
Finally, the fidelity results
\begin{equation}\label{Ftwo}
{\cal F}=\frac{1}{8}\sum_{i=0}^3 F^{i\,i}_{i\,i}+
\frac{1}{24}\sum_{i \ne j=0}^3 \left(F^{i\,i}_{j\,j}
+F^{j\,i}_{i\,j}\right)
\,.
\end{equation}
In this case ${\cal F}$ does not depend on time since 
the gate operation is realized 
with the above definite steps.

\section{Bounds for quantum computation}

If one accounts 
the recent breakthroughs in the real of fault tolerant quantum 
computation \cite{FAULT},
an arbitrarily large quantum computations can be 
performed accurately
provided to have an high degree of accuracy on the 
single gate operation.
This sets an obvious figure of merit for quantum 
computation technology, namely, 
the expected probability of error in one quantum 
gate, which should be of the order of $10^{-6}$
\cite{LLOYD}.

In Fig.1 we show the deviation from the perfect gate fidelity 
as a function of the fractional error of the pulse area. 
In particular the lower straight line concerns the one-bit 
gate for a $\pi$ rotation (\ref{Fone}), 
and the upper straight line the universal 
two-bit gate (\ref{Ftwo}). 
It is known \cite{BOTO} that the value of
$\tau\approx 10^{-8}$ ${\rm s}$ gives the best fit for 
the experimental data of Ref.\cite{WIN},
i.e. $\Omega\approx 10^5$ ${\rm s}^{-1}$, 
$\eta\approx 10^{-1}$.
It means to have $\Omega\tau\approx 10^{-3}$
whose corresponding value of accuracy, i.e.
$1-{\cal F}$, is quite far from the desired one, 
i.e. $10^{-6}$.
In the case of universal two-bit gate, 
we have a better value of accuracy (upper straight line)
already for $\Omega'\tau$ slightly 
less than $\Omega\tau$. 
It turns out that the main 
limitations rely on the single bit
rotation instead on universal two-bit gate.
This fact, though counterintuitive,
can be easily understood if one consider the expression
(\ref{statA}) for the two cases.
It results $\sigma^2(A)/\sigma^2(A')
=\sqrt{N_a}/\eta>1$ for $\pi$-pulses.
This means that the coupling of the 
qubits with the vibrational degree of freedom
makes the two-bit gate less affected 
by the noise with respect to the one-bit gate.
On the other hand, the latter is built up with
on resonance pulse, hence it is more sensitive
to noisy effects,
differently from the two-bit gate
where off resonant pulses are used.
However, since actually 
$\eta/\sqrt{N_a}\approx 10^{-1}$, 
also in the two-bit gate the accuracy falls very far 
from the desirable value. 

Based on these results, we can  
state that quantum information processing 
on a large scale is unrealistic.
As a matter of fact, the fault tolerant quantum 
computation requires a value $\Omega\tau\approx
10^{-6}$, i.e. $\tau\approx 10^{-11}$ ${\rm s}$ 
for $\Omega\approx 10^5$ ${\rm s}^{-1}$.
Since the finite value of $\tau$ is related to 
the Rabi frequency fluctuations
(or in turn to laser intensity fluctuations) \cite{BOTO}, 
this means to improve the laser stability by a 
factor $10^2$ at least!

It is also to remark that, 
within the presented non-dissipative decoherence theory,
the linear behavior 
of the gate fidelity (Fig.1) is typical
of the limit $\Omega\tau\ll 1$.

A rough estimation of the capabilities of a quantum computer
in the presence of non-dissipative decoherence 
can be also made with the following arguments.
Let us consider the optimistic case of
a single run of the Shor's algorithm, and the use of the universal 
two-bit gate operations.
Then, the non-dissipative decoherence theory \cite{RB} shows 
that in the limit of
$\Omega'\tau\ll 1$ the decay rate can be written as 
$\gamma=2\Omega'^2\tau$ (see e.g. Eq.(\ref{ga})). 
Let us now suppose to factorize a $L$-bit number. 
Then, the number of trapped ions $N_a$ should be 
at least $5L$ \cite{PLENIO}.
The time needed for a single run of the Shor's 
algorithm is given by the time required for an
elementary logical operation $4\pi\sqrt{5L}/\eta\Omega$ 
multiplied by the required number of
elementary operations, approximately $(10\,L)^3$ \cite{PLENIO}.  
Of course this product should be
much less then the docoherence time $\gamma^{-1}$. 
Nevertheless, by using the actual experimental values
for parameters (and the corresponding value of 
$\tau\approx 10^{-8}$), 
it results clear that the
factorization of even a four-bit number results impossible.

\section{Conclusion}

In conclusion we have studied the limitations imposed by 
non-dissipative decoherence on quantum computation.
Practically, we have seen that non-dissipative decoherence 
actually constitute a serious impediment
to realize quantum computer beside dissipative decoherence.
We have used a model able to accurately describe the 
decoherence phenomena on ion trap based experiments
caused by the fluctuations of classical quantities.
We have shown that large scale computation seems impossible 
with the present proposals. 
Our results indicate that even a computationally modest goal 
will be extremely
challenging experimentally \cite{HAR}.

Although the conclusions of this paper are rather pessimistic
with regard to the practical application of quantum computers 
for actual computation, there are applications requiring much fewer
operations which are worth considering \cite{NIST}.

\section*{Acknowledgments}

We would like to thank D. Vitali for helpful comments.

\section*{Appendix I}

In this Appendix we explicitly calculate 
the tensor elements (\ref{Ftens}) 
for one-bit gate which are useful to 
calculate the gate fidelity.
We make use of Eqs.(\ref{Rbar}), 
(\ref{Ftens}) and (\ref{Ug}), (\ref{Ue}) obtaining
\begin{eqnarray}
F^{0\,0}_{0\,0}&=&
F^{1\,1}_{1\,1}=F^{1\,0}_{0\,1}=F^{0\,1}_{1\,0}=
\int_0^{\infty} \, dA \, P(t,A) \cos^2[(A-\Omega t)/2]\nonumber\\
&=&\frac{1}{2(1+\Omega^2\tau^2)^{t/2\tau}}
\left\{(1+\Omega^2\tau^2)^{t/2\tau}
+\cos(\Omega t)\cos\left[\frac{t\arctan(\Omega\tau)}{\tau}\right]
+\sin(\Omega t)\sin\left[\frac{t\arctan(\Omega\tau)}{\tau}\right]
\right\}
\,,\\
F^{1\,1}_{0\,0}&=&F^{0\,0}_{1\,1}=\left(1-F^{0\,0}_{0\,0}\right)\,.
\end{eqnarray}
The obtained results are valid in any regime even if  
we are interested in $\Omega\tau\ll 1$ and $t=\pi/\Omega$.

\section*{Appendix II}

In this Appendix we explicitly calculate the 
tensor elements (\ref{Ftens}) 
for two-bit gate which are useful to calculate the gate fidelity.
Let us first examine the effects of evolution operators 
(\ref{UI}), (\ref{UII}), (\ref{UIII})
on the two-bit vector basis.
For $i=0$, we have
\begin{equation}
U_{III}U_{II}U_{I}|g\rangle_1|g\rangle_2|0\rangle
=|g\rangle_1|g\rangle_2|0\rangle\,.
\end{equation}
For $i=1$, we have
\begin{equation}
U_{III}U_{II}U_{I}|g\rangle_1|e\rangle_2|0\rangle
=|g\rangle_1|e\rangle_2|0\rangle\,.
\end{equation}
For $i=2$, we have
\begin{eqnarray}
U_{III}U_{II}U_{I}|e\rangle_1|g\rangle_2|0\rangle
&=&\cos\left[\Omega't_1/2\right]
\left\{ \cos\left[\Omega't_3/2\right]  
|e\rangle_1|g\rangle_2|0\rangle
-i \sin\left[\Omega't_3/2\right]  
|g\rangle_1|g\rangle_2|1\rangle \right\}
\nonumber\\
&&-i\sin\left[\Omega't_1/2\right] 
\cos\left[\Omega't_2/2\right]
\left\{ \cos\left[\Omega't_3/2\right]  
|g\rangle_1|g\rangle_2|1\rangle
-i \sin\left[\Omega't_3/2\right]  
|e\rangle_1|g\rangle_2|0\rangle \right\}
\nonumber\\
&&-\sin\left[\Omega't_1/2\right] 
\sin\left[\Omega't_2/2\right]
|g\rangle_1 |e'\rangle_2 |0\rangle
\,.
\end{eqnarray}
For $i=3$, we have
\begin{eqnarray}
U_{III}U_{II}U_{I}|e\rangle_1|e\rangle_2|0\rangle
&=&\cos\left[\Omega't_1/2\right]
\left\{ \cos\left[\Omega't_3/2\right]  
|e\rangle_1|e\rangle_2|0\rangle
-i \sin\left[\Omega't_3/2\right]  
|g\rangle_1|e\rangle_2|1\rangle \right\}
\nonumber\\
&&-i\sin\left[\Omega't_1/2\right] 
\left\{ \cos\left[\Omega't_3/2\right]  
|g\rangle_1|e\rangle_2|1\rangle
-i \sin\left[\Omega't_3/2\right]  
|e\rangle_1|e\rangle_2|0\rangle \right\}
\,.
\end{eqnarray}
Then, by using the above results in
Eqs.(\ref{Rbar}) and (\ref{Ftens}), we get 
\begin{eqnarray}
F^{0\,0}_{0\,0}&=&F^{1\,1}_{1\,1}
=F^{1\,0}_{0\,1}=F^{0\,1}_{1\,0}=1
\,,\\
F^{2\,2}_{2\,2}&=&\left[C_2(\pi/\Omega')\right]^2
+\left[S_2(\pi/\Omega')\right]^2 C_2(2\pi/\Omega')
-2\left[Z(\pi/\Omega')\right]^2 C_1(2\pi/\Omega')
\,,\\
F^{3\,3}_{3\,3}&=&\left[C_2(\pi/\Omega')\right]^2
+\left[S_2(\pi/\Omega')\right]^2
-2\left[Z(\pi/\Omega')\right]^2
\,,\\
F^{2\,0}_{0\,2}&=&F^{2\,1}_{1\,2}
=F^{0\,2}_{2\,0}=F^{1\,2}_{2\,1}
=\left[C_1(\pi/\Omega')\right]^2
-\left[S_1(\pi/\Omega')\right]^2 C_1(2\pi/\Omega')
\,,\\
F^{3\,0}_{0\,3}&=&F^{3\,1}_{1\,3}=F^{0\,3}_{3\,0}=F^{1\,3}_{3\,1}
=-\left[C_1(\pi/\Omega')\right]^2
+\left[S_1(\pi/\Omega')\right]^2
\,,\\
F^{3\,2}_{2\,3}&=&F^{2\,3}_{3\,2}
=-\left[C_2(\pi/\Omega')\right]^2
-\left[S_2(\pi/\Omega')\right]^2 C_1(2\pi/\Omega')
+\left[Z(\pi/\Omega')\right]^2
+\left[Z(\pi/\Omega')\right]^2 C_1(2\pi/\Omega')
\,,\\
F^{0\,0}_{1\,1}&=&F^{0\,0}_{2\,2}=F^{0\,0}_{3\,3}=
F^{1\,1}_{0\,0}=F^{1\,1}_{2\,2}=F^{1\,1}_{3\,3}=
F^{2\,2}_{0\,0}=F^{2\,2}_{1\,1}=F^{2\,2}_{3\,3}=
F^{3\,3}_{0\,0}=F^{3\,3}_{1\,1}=F^{3\,3}_{2\,3}=0
\,,
\end{eqnarray}
where
\begin{eqnarray}
C_2(t)&=&\int_0^{\infty} \, dA' \, P(t,A') 
\cos^2[A'/2]
=\frac{1}{2(1+\Omega'^2\tau^2)^{t/2\tau}}
\left\{(1+\Omega'^2\tau^2)^{t/2\tau}
+\cos\left[\frac{t\arctan(\Omega'\tau)}{\tau}\right]
\right\}\,,
\\
Z(t)&=&\int_0^{\infty} \, dA' \, P(t,A') 
\sin[A'/2] \cos[A'/2]
=\frac{1}{2(1+\Omega'^2\tau^2)^{t/2\tau}}\left\{
\sin\left[\frac{t\arctan(\Omega'\tau)}{\tau}\right]
\right\}\,,
\\
S_2(t)&=&\int_0^{\infty} \, dA' \, P(t,A') \sin^2[A'/2]
=1-C_2(t)\,,
\\
C_1(t)&=&\int_0^{\infty} \, dA' \, P(t,A') \cos[A'/2]
=\frac{1}{(1+\Omega'^2\tau^2/4)^{t/2\tau}}\left\{
\cos\left[\frac{t\arctan(\Omega'\tau/2)}{\tau}\right]
\right\}\,,
\\
S_1(t)&=&\int_0^{\infty} \, dA' \, P(t,A') \sin[A'/2]
=\frac{1}{(1+\Omega'^2\tau^2/4)^{t/2\tau}}\left\{
\sin\left[\frac{t\arctan(\Omega'\tau/2)}{\tau}\right]
\right\}\,.
\end{eqnarray}
The obtained results are valid in any regime even if  
we are interested in $\Omega'\tau\ll 1$.

\bibliographystyle{unsrt}

\begin{figure}[htb]

\centerline{\epsfig{figure=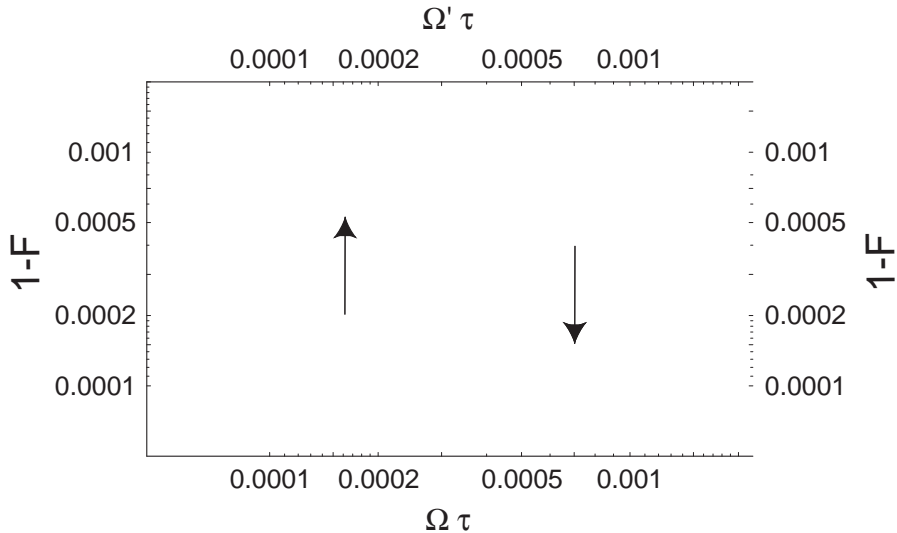,width=12cm}}

\caption{
The quantity $1-{\cal F}$ vs the fractional error 
of the pulse area is shown in a $\log$-$\log$ plot. 
The lower straight line concerns a one-bit gate 
(after a $\pi$ rotation).
The upper straight line concerns the universal two-bit gate.}

\end{figure}

\end{document}